\documentstyle[12pt,graphicx]{article}

\def\la{\mathrel{\mathpalette\fun <}}
\def\ga{\mathrel{\mathpalette\fun >}}
\def\fun#1#2{\lower3.6pt\vbox{\baselineskip0pt\lineskip.9pt
  \ialign{$\mathsurround=0pt#1\hfil##\hfil$\crcr#2\crcr\sim\crcr}}}

\begin{document}
\title{Probing Physics at Extreme Energies with Cosmic Ultra-High
Energy Radiation}

\author{G\"unter Sigl\\
GReCO, Institut d'Astrophysique de Paris, CNRS, 98bis Boulevard Arago,\\
75014 Paris, France}

\maketitle

\abstract{
The highest energy cosmic rays observed possess macroscopic energies and
their origin is likely to be associated with the most energetic processes
in the Universe. Their existence triggered a flurry of theoretical
explanations ranging from conventional shock acceleration to particle
physics beyond the Standard Model and processes taking place at the
earliest moments of our Universe. Furthermore, many new experimental
activities promise a strong increase of statistics at the highest
energies and a combination with $\gamma-$ray and neutrino astrophysics
will put strong constraints on these theoretical models.
We give an overview over this quickly evolving research field with
focus on testing new particle physics.}

\section{Introduction}

Over the last few years, several giant air showers have been
detected confirming the arrival of cosmic rays (CRs) with energies
up to a few hundred EeV (1 EeV $\equiv 10^{18}$
eV)~\cite{haverah,fe,agasa}. The existence of such ultra-high energy
cosmic rays (UHECRs)
pose a serious challenge for conventional theories of
CR origin based on acceleration of charged particles in powerful
astrophysical objects. The question of the origin of these UHECRs
is, therefore, currently a subject of much intense debate and
discussions as well as experimental efforts; see
Ref.~\cite{reviews} for recent brief reviews, and Ref.~\cite{bs-rev}
for a detailed review.

The problems encountered in trying to explain UHECRs in terms of
``bottom-up'' acceleration mechanisms have been well-documented
in a number of studies;
see, e.g., Refs.~\cite{hillas-araa,ssb,norman}. It is hard to accelerate 
protons and heavy nuclei up to such energies even in the most
powerful astrophysical objects such as radio
galaxies and active galactic nuclei. Also, nucleons above
$\simeq70\,$EeV lose energy drastically due to 
photo-pion production on the cosmic microwave background (CMB) 
--- the Greisen-Zatsepin-Kuzmin (GZK) effect~\cite{gzk} --- 
which limits the distance to possible sources to less than
$\simeq100\,$Mpc~\cite{ssb}. Heavy nuclei are photodisintegrated
in the CMB within a few Mpc~\cite{heavy}. There are no 
obvious astronomical sources within $\simeq 100$ Mpc of the
Earth~\cite{elb-som,ssb}.

The distance restriction imposed by the GZK effect can be circumvented if
the problem of energetics is somehow solved separately and if one
postulates new particles beyond the Standard Model; this will
be discussed in the third section.

In contrast, in the ``top-down'' scenarios, which will be discussed
in the last section, the problem of energetics is
trivially solved. Here, the UHECR particles are the
decay products of some supermassive ``X'' particles of mass
$m_X\gg10^{20}\,$eV, and have energies all the way up to $\sim m_X$. Thus,
no acceleration mechanism is needed. The massive X
particles could be metastable relics of the early Universe
with lifetimes of the order the current age of the Universe
or could be released from topological defects
that were produced in the early Universe during 
symmetry-breaking phase transitions envisaged in Grand Unified
Theories (GUTs). If the X particles
themselves or their sources cluster similar to dark matter,
the dominant observable  UHECR contribution would come from the
Galactic Halo and absorption would be negligible.

The main problem of non-astrophysical solutions of the UHECR
problem in general is that they are highly model dependent.
On the other hand, they allow to at least test new physics beyond
the Standard Model of particle physics
(such as Grand Unification and new interactions beyond the reach of
terrestrial accelerators) as well as early Universe cosmology
(such as topological defects and/or massive particle production in
inflation) at energies often inaccessible to accelerator experiments.

The physics and astrophysics of UHECRs are intimately linked with
the emerging field of neutrino astronomy (for reviews see
Refs.~\cite{ghs,halzen}) as well as with the already
established field of $\gamma-$ray astronomy (for reviews see, e.g.,
Ref.~\cite{gammarev}). Indeed, all
scenarios of UHECR origin, including the top-down models, are severely
constrained by neutrino and $\gamma-$ray observations and limits.
In turn, this linkage has important consequences for theoretical
predictions of fluxes of extragalactic neutrinos above a TeV
or so whose detection is a major goal of next-generation
neutrino telescopes: If these neutrinos are
produced as secondaries of protons accelerated in astrophysical
sources and if these protons are not absorbed in the sources,
but rather contribute to the UHECR flux observed, then
the energy content in the neutrino flux can not be higher
than the one in UHECRs, leading to the so called Waxman Bahcall
bound for sources with soft acceleration spectra~\cite{wb-bound,mpr}.
If one of these assumptions does not apply, such as for acceleration
sources with injection spectra harder than $E^{-2}$ and/or opaque
to nucleons, or in the top-down scenarios where X particle decays
produce much fewer nucleons than $\gamma-$rays and neutrinos,
the Waxman Bahcall bound does not apply, but the neutrino
flux is still constrained by the observed diffuse $\gamma-$ray
flux in the GeV range.

\section{Propagation of Ultra-High Energy Radiation: Simulations}
Before discussing specific scenarios for UHECR origin we
give a short account of the numerical tools used to compute
spectra of ultra-high energy cosmic and $\gamma-$rays, and
neutrinos~\cite{code,kkss1,kkss2}. In the following we assume a
flat Universe with a Hubble constant of
$H=70\;{\rm km}\;{\rm sec}^{-1}{\rm Mpc}^{-1}$ and a cosmological
constant $\Omega_\Lambda=0.7$, as favored by current observations.

The relevant nucleon interactions implemented are
pair production by protons ($p\gamma_b\rightarrow p e^- e^+$),
photoproduction of single or multiple pions ($N\gamma_b \rightarrow N
\;n\pi$, $n\geq1$), and neutron decay.

$\gamma$-rays and electrons/positrons initiate  electromagnetic
(EM) cascades on low energy radiation fields such as the
CMB. The high energy photons undergo electron-positron pair
production (PP; $\gamma \gamma_b \rightarrow e^- e^+$), and 
at energies below $\sim 10^{14}$ eV they interact mainly with 
the universal infrared and optical (IR/O) backgrounds, while above 
$\sim 100$ EeV  they interact mainly with the universal radio background (URB).
In the Klein-Nishina regime, where the CM energy is
large compared to the electron mass, one of the outgoing particles usually
carries most of the initial energy. This ``leading''
electron (positron) in turn can transfer almost all of its energy to
a background photon via inverse
Compton scattering (ICS; $e \gamma_b \rightarrow e^\prime\gamma$).
EM cascades are driven by this cycle of PP and ICS.
The energy degradation of the ``leading'' particle in this cycle
is slow, whereas the total number of particles grows
exponentially with time. All
EM interactions that influence the $\gamma$-ray spectrum in the energy range
$10^8\,{\rm eV} < E < 10^{25}\,$eV, namely PP, ICS, triplet pair
production (TPP; $e \gamma_b
\rightarrow e e^- e^+$), and double pair production (DPP, $\gamma \gamma_b
\rightarrow e^-e^+e^-e^+$), as well as synchrotron losses
of electrons in the large scale extragalactic magnetic field
(EGMF), are included.

Similarly to photons, UHE neutrinos give rise to neutrino
cascades in the primordial neutrino background via exchange
of W and Z bosons~\cite{zburst1}. Besides the secondary
neutrinos which drive the neutrino cascade, the W and Z decay products
include charged leptons and quarks which in turn feed into the
EM and hadronic channels. Neutrino interactions become
especially significant if the relic neutrinos have masses $m_\nu$
in the eV range and thus constitute hot dark matter, because
the Z boson resonance then occurs at an UHE neutrino energy
$E_{\rm res}=4\times10^{21}({\rm eV}/m_\nu)$ eV. In fact, the
decay products of this ``Z-burst'' have been
proposed as a significant source of UHECRs~\cite{zburst2}.
The big drawback of this scenario is the need of enormous
primary neutrino fluxes that cannot be produced by known astrophysical
acceleration sources~\cite{kkss1}, and thus most likely
requires a more exotic top-down type source such as X particles
exclusively decaying into neutrinos~\cite{gk}. Even this possibility
appears close to being ruled out due to a tendency to overproduce
the diffuse GeV $\gamma-$ray flux observed by EGRET~\cite{egret,kkss2}

The two major uncertainties in the particle transport are the
intensity and spectrum of the URB for which there exist
no direct measurements in the relevant MHz regime~\cite{Clark,PB},
and the average value of the EGMF. Simulations
have been performed for different assumptions on these.
A strong URB tends
to suppress the UHE $\gamma$-ray flux by direct absorption
whereas a strong EGMF blocks EM cascading (which otherwise develops
efficiently especially in a low URB) by synchrotron cooling
of the electrons. For the IR/O background we used the most
recent data~\cite{irb}.

In top-down scenarios, the particle injection spectrum is generally dominated
by the ``primary'' $\gamma$-rays and neutrinos over nucleons. These
primary $\gamma$-rays and neutrinos are produced by the decay of
the primary pions resulting from the hadronization of quarks that come
from the decay of the X particles. In contrast, in acceleration scenarios
the primaries are accelerated protons or nuclei, and
$\gamma$-rays, electrons, and neutrinos are produced as secondaries
from decaying pions that are in turn produced by the interactions
of nucleons with the CMB.

\section{New Primary Particles and Interactions}

A possible way around the problem of missing counterparts
within acceleration scenarios is to propose primary
particles whose range is not limited by interactions with
the CMB. Within the Standard Model the only candidate is the neutrino,
whereas in extensions of the Standard Model one could think of
new neutrals such as axions or stable supersymmetric elementary
particles. Such options are mostly ruled out by the tension
between enforcing small EM coupling and large hadronic coupling
to ensure normal air showers~\cite{ggs}. Also suggested have been
new neutral hadronic bound states of light gluinos with
quarks and gluons, so-called R-hadrons that are heavier than
nucleons, and therefore have a higher GZK threshold~\cite{cfk}.
Since this too seems to be disfavored by accelerator
constraints~\cite{gluino} we will here focus on neutrinos.

In both the neutrino and new neutral stable particle scenario
the particle propagating over extragalactic distances would have
to be produced as a secondary in
interactions of a primary proton that is accelerated in
a powerful AGN which can, in contrast to the case of
extensive air showers (EAS) induced by nucleons, nuclei, or $\gamma-$rays,
be located at high redshift. Consequently, these scenarios predict
a correlation between primary arrival directions
and high redshift sources. In fact, possible evidence
for a correlation of UHECR arrival directions with compact radio
quasars and BL-Lac objects, some of them possibly too far away to
be consistent with the GZK effect,
was recently reported~\cite{bllac}. Only a few more events could confirm or
rule out the correlation hypothesis. Note, however, that
these scenarios require the primary proton to be accelerated
up to at least $10^{21}\,$eV, demanding a very powerful
astrophysical accelerator.

\subsection{New Neutrino Interactions}

Neutrino primaries have the advantage of being well established
particles. However, within the Standard Model their interaction cross
section with nucleons, whose charged current part can
be parametrized by~\cite{gqrs}
\begin{equation}
\sigma^{SM}_{\nu N}(E)\simeq2.36\times10^{-32}(E/10^{19}
  \,{\rm eV})^{0.363}\,{\rm cm}^2\,,\label{cross}
\end{equation}
for $10^{16}\,{\rm eV}\la E\la10^{21}\,{\rm eV}$,
falls short by about five orders of
magnitude to produce ordinary air showers.
However, it has been suggested that the neutrino-nucleon
cross section, $\sigma_{\nu N}$, can be enhanced by new
physics beyond the electroweak scale in the center of
mass (CM) frame, or above about a PeV in the nucleon rest frame.
Neutrino induced air showers may therefore rather directly
probe new physics beyond the electroweak scale.

One possibility consists of a large increase
in the number of degrees of freedom above the electroweak 
scale~\cite{kovesi-domokos}. A specific implementation
of this idea is given in theories with $n$ additional large
compact dimensions and a quantum gravity scale $M_{4+n}\sim\,$TeV
that has recently received much attention in the
literature~\cite{tev-qg} because it provides an alternative
solution (i.e., without supersymmetry) to the hierarchy problem
in grand unifications of gauge interactions. It turns
out that the largest contribution to the neutrino-nucleon
cross section is provided by the production of microscopic
black holes centered on the brane representing our world,
but extending into the extra dimensions. The cross sections
can be larger than the Standard model one by up to a factor
100~\cite{fs}. This is not sufficient to explain the observed
UHECR events~\cite{kp}.

\begin{figure}[htb]
\includegraphics[width=0.7\textwidth,clip=true,angle=270]{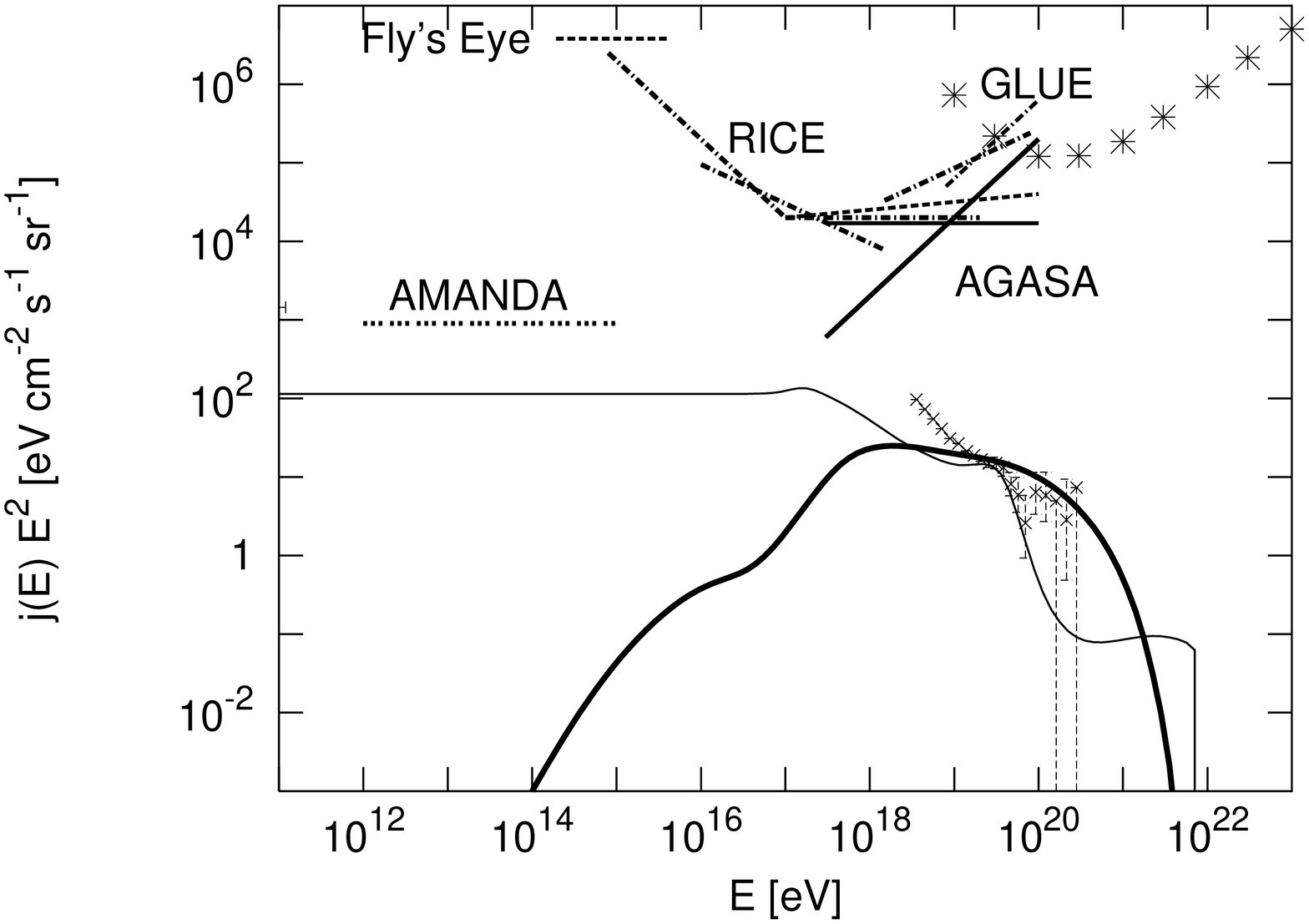}
\caption[...]{(from work in Ref.~\cite{kkss2}) Cosmogenic neutrino
flux per flavor (thick line,
assuming maximal mixing among all flavors) from primary proton
flux (thin line) fitted to the AGASA cosmic ray data~\cite{agasa}
above $3\times10^{18}\,$eV (error bars). The UHECR sources were assumed to
inject a $E^{-2}$ proton spectrum up to $10^{22}\,$eV with
luminosity $\propto(1+z)^3$ up to
$z=2$. Also shown are existing upper limits on the diffuse neutrino fluxes
from AMANDA~\cite{amanda_limit} AGASA~\cite{agasa_nu}, the
Fly's Eye~\cite{baltrusaitis} and RICE~\cite{rice} experiments, and
the limit obtained with the Goldstone radio telescope (GLUE)~\cite{glue}.}
\label{fig1}
\end{figure}

However, the UHECR data can be used to put constraints on cross sections
satisfying $\sigma_{\nu N}(E\ga10^{19}\,{\rm eV})
\la10^{-27}\,{\rm cm}^2$. Particles with such cross
sections would give rise to horizontal air showers which have
not yet been observed. Resulting upper limits on their fluxes
assuming the Standard Model cross section Eq.~(\ref{cross}) are
shown in Fig.~\ref{fig1}. Comparison with the ``cosmogenic'' neutrino flux
produced by UHECRs interacting with the CMB then results in upper
limits on the cross section which are about a factor 1000 larger
than Eq.~(\ref{cross}) in the energy range between $\simeq10^{17}\,$eV
and $\simeq10^{19}\,$eV~\cite{tol}. The projected sensitivity of
future experiments shown in Fig.~\ref{fig2} below indicate that these limits
could be lowered down to the Standard Model one~\cite{afgs}. In case
of a detection of penetrating events the degeneracy of the cross
section with the unknown flux could be broken by comparing horizontal
air showers with Earth skimming events~\cite{kw}.

\section{Top-Down Scenarios}

As mentioned in the introduction, all top-down scenarios
involve the decay of X particles of mass close to the GUT scale
which can basically be produced in two ways: If they are very
short lived, as usually expected in many GUTs, they have to be
produced continuously. The only way this can be achieved is
by emission from topological defects left over from cosmological
phase transitions that may have occurred in the early Universe at
temperatures close to the GUT scale, possibly during reheating
after inflation. Topological defects
necessarily occur between regions that are causally disconnected, such
that the orientation of the order parameter
associated with the phase transition, can not be communicated
between these regions and consequently will adopt different
values. Examples are cosmic strings, magnetic monopoles, and
domain walls. The defect density is consequently given by the particle horizon
in the early Universe.
The defects are topologically stable, but time dependent motion
leads to the emission of particles with a mass comparable to the
temperature at which the phase transition took place. The
associated phase transition can also occur during reheating
after inflation.

Alternatively, instead of being released from topological
defects, X particles
may have been produced directly in the early Universe and,
due to some unknown symmetries, have a very
long lifetime comparable to the age of the Universe.
In contrast to Weakly-Interacting Massive Particles (WIMPS)
below a few hundred TeV which are the usual dark matter
candidates motivated by, for example, supersymmetry and can
be produced by thermal freeze out, such superheavy X particles
have to be produced non-thermally (see Ref.~\cite{kuz-tak}
for a review). In all these cases, such particles,
also called ``WIMPZILLAs'', would contribute to the dark matter
and their decays could still contribute to UHECR fluxes today,
with an anisotropy pattern that reflects the dark matter
distribution in the halo of our Galaxy.

It is interesting to note that one of the prime motivations
of the inflationary paradigm was to dilute excessive production
of ``dangerous relics'' such as topological defects and
superheavy stable particles. However, such objects can be
produced right after inflation during reheating
in cosmologically interesting abundances, and with a mass scale
roughly given by the inflationary scale which in turn
is fixed by the CMB anisotropies to
$\sim10^{13}\,$GeV~\cite{kuz-tak}. The reader will realize that
this mass scale is somewhat above the highest energies
observed in CRs, which implies that the decay products of
these primordial relics could well have something to do with
UHECRs which in turn can probe such scenarios!

For dimensional reasons the spatially averaged X particle
injection rate can only
depend on the mass scale $m_X$ and on cosmic time $t$ in the
combination
\begin{equation}
  \dot n_X(t)=\kappa m_X^p t^{-4+p}\,,\label{dotnx}
\end{equation}
where $\kappa$ and $p$ are dimensionless constants whose
value depend on the specific top-down scenario~\cite{BHS},
For example, the case $p=1$ is representative of scenarios
involving release of X particles from topological defects,
such as ordinary cosmic
strings~\cite{BR}, necklaces~\cite{BV} and magnetic 
monopoles~\cite{BS}. This can be easily seen as follows:
The energy density $\rho_s$ in a network of defects has to scale
roughly as the critical density, $\rho_s\propto\rho_{\rm crit}\propto
t^{-2}$, where $t$ is cosmic time, otherwise the defects
would either start to overclose the Universe, or end up
having a negligible contribution to the total energy
density. In order to maintain this scaling, the defect
network has to release energy with a rate given by
$\dot\rho_s=-a\rho_s/t\propto t^{-3}$, where $a=1$ in
the radiation dominated aera, and $a=2/3$ during matter
domination. If most of this energy goes into emission
of X particles, then typically $\kappa\sim{\cal O}(1)$.
In the numerical simulations presented below, it was
assumed that the X particles are nonrelativistic at decay.

The X particles could be gauge bosons, Higgs bosons, superheavy fermions,
etc.~depending on the specific GUT. They would have
a mass $m_X$ comparable to the symmetry breaking scale and would
decay into leptons and/or quarks of roughly
comparable energy. The quarks interact strongly and 
hadronize into nucleons ($N$s) and pions, the latter
decaying in turn into $\gamma$-rays, electrons, and neutrinos. 
Given the X particle production rate, $dn_X/dt$, the effective
injection spectrum of particle species $a$ ($a=\gamma,N,e^\pm,\nu$) 
via the hadronic channel can be
written as $(dn_X/dt)(2/m_X)(dN_a/dx)$,
where $x \equiv 2E/m_X$, and $dN_a/dx$ is the relevant
fragmentation function (FF).

We adopt the Local Parton Hadron Duality (LPHD) approximation~\cite{detal}
according to which the total
hadronic FF, $dN_h/dx$, is taken to be proportional to the spectrum
of the partons (quarks/gluons) in the parton cascade (which is initiated
by the quark through perturbative QCD processes) after evolving the parton
cascade to a stage where the typical transverse momentum transfer in the
QCD cascading processes has come down to $\sim R^{-1}\sim$ few hundred 
MeV, where $R$ is a typical hadron size. The parton spectrum is obtained
from solutions of the standard QCD evolution equations in modified leading
logarithmic approximation (MLLA) which provides good fits to accelerator
data at LEP energies~\cite{detal}. Within the LPHD hypothesis, the pions
and nucleons after hadronization have essentially the same spectrum. 
The LPHD does not, however, fix the relative abundance of pions and
nucleons after hadronization. Motivated by accelerator data, we assume
the nucleon content $f_N$ of the hadrons to be $\simeq10$\%, and the
rest pions distributed equally among the three charge states.
Recent work suggests that the nucleon-to-pion ratio may be significantly
higher in certain
ranges of $x$ values at the extremely high energies of interest
here~\cite{frag}, but the situation is not completely
settled yet.

\subsection{Predicted Fluxes}

\begin{figure}[htb]
\includegraphics[width=0.7\textwidth,clip=true,angle=270]{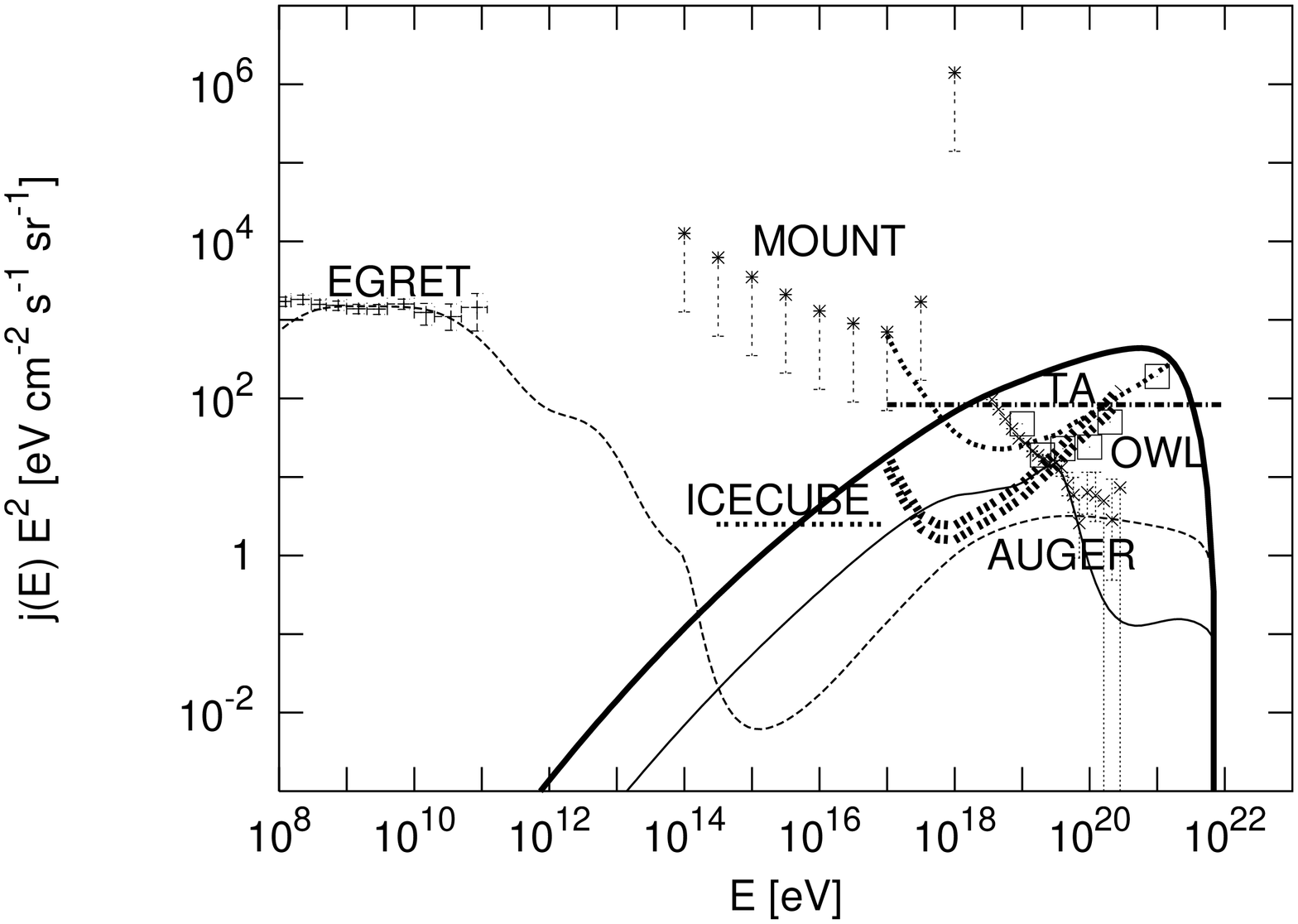}
\caption[...]{(from work in Ref.~\cite{kkss2}) Predictions for the
differential fluxes of
$\gamma-$rays (dotted line), nucleons (thin solid line), and neutrinos
per flavor (thick solid line, assuming maximal mixing among all flavors)
in a TD model characterized by $p=1$, $m_X = 2\times10^{14}\,$GeV,
and the decay mode $X\to q+q$, assuming the QCD fragmentation function
in MLLA approximation~\cite{detal}, with a fraction of
10\% nucleons. The calculation used the code described in Ref.~\cite{code}
and assumed the minimal URB version consistent with observations~\cite{PB}
and an EGMF of $10^{-12}\,$G. Cosmic ray data are as in Fig.~\ref{fig1}
and the EGRET data on the left margin represents the
diffuse $\gamma-$ray flux between 30 MeV and 100 GeV~\cite{egret}.
Also shown are expected sensitivities of the Auger
project currently in construction to electron/muon and
tau-neutrinos~\cite{auger_nu}, and the planned projects telescope
array~\cite{ta_nu}, the water-based ANTARES~\cite{antares},
the ice-based ICECUBE~\cite{icecube}, the
fluorescence/\v{C}erenkov detector MOUNT~\cite{mount}, and the
space based OWL~\cite{owl_nu} (we take the latter as representative
also for EUSO~\cite{euso}).
\label{fig2}}
\end{figure}

Fig.~\ref{fig2} shows results for the time averaged nucleon, $\gamma-$ray,
and neutrino fluxes in a typical TD scenario, along with low energy
$\gamma-$ray flux constraints and neutrino flux sensitivities of
future experiments. The spectrum was optimally normalized 
to allow for an explanation of the observed UHECR
events, assuming their consistency with a nucleon or
$\gamma-$ray primary. The flux below $\la2\times10^{19}\,$eV
is presumably due to conventional
acceleration in astrophysical sources and was not fit.
The PP process on the CMB depletes the photon flux above 100 TeV, and the
same process on the IR/O background causes depletion of the photon flux in
the range 100 GeV--100 TeV, recycling the absorbed energies to
energies below 100 GeV through EM cascading.
The predicted background is {\it not} very sensitive to
the specific IR/O background model, however~\cite{ahacoppi}.
The scenario in Fig.~\ref{fig2} obviously
obeys all current constraints within the normalization
ambiguities and is therefore quite viable. Note
that the diffuse $\gamma-$ray background measured by
EGRET~\cite{egret} up to 10 GeV puts a strong constraint on these
scenarios, especially if there is already a significant
contribution to this background from conventional
sources such as unresolved $\gamma-$ray blazars~\cite{muk-chiang}.
However, this constraint is much weaker for TDs or decaying long
lived X particles with a non-uniform clustered density~\cite{bkv}

The energy loss and absorption lengths for UHE nucleons and photons
are short ($\la100$ Mpc). Thus, their predicted UHE fluxes are
independent of cosmological evolution. The $\gamma-$ray flux
below $\simeq10^{11}\,$eV, however, scales as the
total X particle energy release integrated over all redshifts
and increases with decreasing $p$~\cite{sjsb}. For
$m_X=2\times10^{16}\,$GeV, scenarios with $p<1$ are therefore ruled
out, whereas constant comoving injection models ($p=2$) are well within the
limits.

It is clear from the above discussions that the predicted particle fluxes
in the TD scenario are currently uncertain to a large extent due to 
particle physics uncertainties (e.g., mass and decay modes of the X
particles, the quark fragmentation function, the nucleon fraction $f_N$,
and so on) as well as astrophysical uncertainties (e.g., strengths of the
radio and infrared backgrounds, extragalactic magnetic fields, etc.). 
More details on the dependence of the predicted UHE particle spectra and
composition on these particle physics and astrophysical
uncertainties are contained in Ref.~\cite{SLBY}.

We stress here that there are viable TD scenarios which
predict nucleon fluxes that are comparable to or even higher than
the $\gamma-$ray flux at all energies, even though $\gamma-$rays
dominate at production.
This occurs, e.g., in the case of high URB
and/or for a strong EGMF, and a nucleon fragmentation fraction of
$\simeq10\%$. Some of these TD 
scenarios would therefore remain viable even if UHECR induced EAS
should be proven inconsistent with photon primaries (see,
e.g., Ref.~\cite{gamma_comp}). This is in contrast to scenarios with
decaying massive dark matter in the Galactic halo which,
due to the lack of absorption, predict compositions directly
given by the fragmentation function, i.e. domination by
$\gamma-$rays.

The normalization procedure to the UHECR flux described above
imposes the constraint $Q^0_{\rm UHECR}\la10^{-22}\,{\rm eV}\,{\rm
cm}^{-3}\,{\rm sec}^{-1}$ within a factor of a
few~\cite{ps1,SLBY,slsc} for the total energy release rate $Q_0$
from TDs at the current epoch.
In most TD models, because of the unknown values of the
parameters involved, it is currently not
possible to calculate the exact value of $Q_0$ from first principles,
although it has been shown that the required values of $Q_0$ (in order to
explain the UHECR flux) mentioned above are quite possible for
certain kinds of TDs. Some cosmic
string simulations and the necklace scenario suggest that
defects may lose most of
their energy in the form of X particles and estimates of this
rate have been given~\cite{vincent,BV}. If that is the case, the
constraint on $Q^0_{\rm UHECR}$ translates via Eq.~(\ref{dotnx})
into a limit on the symmetry
breaking scale $\eta$ and hence on the mass $m_X$ of the X particle: 
$\eta\sim m_X\la10^{13}\,$GeV~\cite{wmgb}. Independently 
of whether or not this scenario explains UHECR, the EGRET measurement
of the diffuse GeV $\gamma-$ray background leads to a similar bound,
$Q^0_{\rm EM}\la2.2\times10^{-23}\,h
(3p-1)\,{\rm eV}\,{\rm cm}^{-3}\,{\rm sec}^{-1}$, which leaves
the bound on $\eta$ and $m_X$ practically unchanged.
Furthermore, constraints from limits on CMB distortions and light
element abundances from $^4$He-photodisintegration are
comparable to the bound from the directly observed
diffuse GeV $\gamma$-rays~\cite{sjsb}. That these crude
normalizations lead to values of $\eta$ in the right range
suggests that defect models require less fine tuning than
decay rates in scenarios of metastable massive dark matter.

As discussed above, in TD scenarios most of the energy is
released in the form of EM particles and neutrinos. If the X
particles decay into a quark and a lepton, the quark hadronizes
mostly into pions and the ratio of energy release into the
neutrino versus EM channel is $r\simeq0.3$. The energy fluence
in neutrinos and $\gamma-$rays is thus comparable. However,
whereas the photons are recycled down to the GeV range where
their flux is constrained by the EGRET measurement, the neutrino
flux is practically not changed during propagation and thus
reflects the injection spectrum. Its predicted level is consistent
with all existing upper limits (compare Fig.~\ref{fig2} with
Fig.~\ref{fig1}) but should be detectable by several experiments
under construction or in the proposal stage (see Fig.~\ref{fig2}).
This would allow to directly see the quark fragmentation spectrum.

\section{Conclusions}

Ultra-high energy cosmic rays have the potential to open a window
to and act as probes of new particle physics beyond the Standard Model
as well as processes occuring in the early Universe at energies close
to the Grand Unification scale. Even if their origin will turn out
to be attributable to astrophysical shock acceleration with no new
physics involved, they will still be witnesses of one of the most energetic
processes in the Universe. The future appears promising and
exciting due to the anticipated arrival of several large scale
experiments.

\end{document}